%% file: main.tex
\documentclass[sigconf]{acmart}
\pdfoutput=1

\AtBeginDocument{%
  \providecommand\BibTeX{{%
    \normalfont B\kern-0.5em{\scshape i\kern-0.25em b}\kern-0.8em\TeX}}}

\usepackage{xspace}

\newcommand{\aka}{\emph{a.k.a.,}\xspace}

\newcommand{\ignore}[1]{}

\usepackage{ragged2e} 
\usepackage{booktabs,makecell, multirow, tabularx}
\usepackage{colortbl}
\usepackage{array}

\newcommand{\model}{GNR\xspace}
\newcommand{\trainsmall}{UIFT\xspace}
\newcommand{\fullmodel}{{Generative News Recommendation}\xspace}

\copyrightyear{2024}
\acmYear{2024}
\setcopyright{rightsretained}
\acmConference[WWW '24]{Proceedings of the ACM Web Conference 2024}{May 13--17, 2024}{Singapore, Singapore}
\acmBooktitle{Proceedings of the ACM Web Conference 2024 (WWW '24), May 13--17, 2024, Singapore, Singapore}\acmDOI{10.1145/3589334.3645448}
\acmISBN{979-8-4007-0171-9/24/05}

\begin{document}            

\title{Generative News Recommendation}

\input{author}

\begin{abstract}
Most existing news recommendation methods tackle this task by conducting semantic matching between candidate news and user representation produced by historical clicked news.
However, they overlook the high-level connections among different news articles and also ignore the profound relationship between these news articles and users.
And the definition of these methods dictates that they can only deliver news articles as-is. 
On the contrary, integrating several relevant news articles into a coherent narrative would assist users in gaining a quicker and more comprehensive understanding of events.
In this paper, we propose a novel generative news recommendation paradigm that includes two steps: (1) Leveraging the internal knowledge and reasoning capabilities of the Large Language Model (LLM) to perform high-level matching between candidate news and user representation; 
(2) Generating a coherent and logically structured narrative based on the associations between related news and user interests, thus engaging users in further reading of the news.
Specifically, we propose \model to implement the generative news recommendation paradigm.
First, we compose the dual-level representation of news and users by leveraging LLM to generate theme-level representations and combine them with semantic-level representations.
Next, in order to generate a coherent narrative, we explore the news relation and filter the related news according to the user preference.
Finally, we propose a novel training method named \trainsmall to train the LLM to fuse multiple news articles in a coherent narrative.
Extensive experiments show that \model can improve recommendation accuracy and eventually generate more personalized and factually consistent narratives~\footnote{We have released our code at https://github.com/morganf33/GNR}.
\end{abstract}



\maketitle
\input{intro.tex}
\input{related.tex}
\input{preliminar.tex}

\input{method.tex}

\input{exp.tex}
\input{conclusion.tex}

\section*{Acknowledgement}
This work was supported by the National Natural Science Foundation of China (T2293773, 62372275, 62272274, 62202271, 62102234, 62072279), the Natural Science Foundation of Shandong Province (ZR2023QF159), the National Key R\&D Program of China with grant No.2022YFC3303004.
\bibliographystyle{ACM-Reference-Format}


\newpage
\appendix
\section{appendix}
\subsection{Example prompts for multi-news narrative generation}\label{sec:fuse_prompt}

\begin{table}[htb]
\small
\renewcommand\arraystretch{1.1}
\begin{tabular}{>{\centering\arraybackslash}p{8.5cm}}
\rowcolor{yellow!20}
\hline
\multicolumn{1}{c}{\textbf{Instruction}}                                                                                                                                                                                                                                                                                                                                                                                                                                                                                                                                                                                                                                                                                                                                                                                                                                                                                                                                                                                                                                                                                                                                                                                                     \\ \hline
\rowcolor{yellow!2}
\begin{tabular}[c]{@{}l@{}}You are a personalized text generator. First, I will provide you with a news \\ list that includes both the \textbf{{[}main news{]}} and \textbf{{[}topic-related news{]}}. Second, \\ I will provide you with user interests, including the \textbf{{[}categories{]}} and \\ \textbf{{[}topics{]}} of news that the user is interested in. Based on the input news list \\ and user interests, you are required to generate a \textbf{\{personalized news} \\ \textbf{summary\}} centered around the \textbf{{[}main news{]}}.\end{tabular}                                                                                                                                                                                                                                                                                                                                                                                                                                                                                                                                                                                                                                                                                                                                                           \\ \hline
\rowcolor{violet!10}
\multicolumn{1}{c}{\textbf{Input}}                                                                                                                                                                                                                                                                                                                                                                                                                                                                                                                                                                                                                                                                                                                                                                                                                                                                                                                                                                                                                                                                                                                                                                                                           \\ \hline

\begin{tabular}[c]{@{}l@{}}News List:\\ \{``ID'': ``Main News'', ``title'': ``Brett Kavanaugh calls Ruth Bader Ginsburg \\ `inspiration,' heaps gratitude on allies'', ``abstract'': `In his first speech as a \\ Supreme Court justice, Brett Kavanaugh heaped ``gratitude'' \\ on his supporters and hailed Ruth Bader Ginsburg as an ``inspiration.'''', \\ ``topic'': ``Brett Kavanaugh, ...''\}\\ \{``ID'': ``N****'', ``title'': ``Ruth Bader Ginsburg misses court due to illness'', \\ ``abstract'': ``Supreme Court Justice Ruth Bader Ginsburg was not on the \\ bench for oral arguments Wednesday due to illness ...'', ``topic'': ``Supreme\\Court, Ruth Bader Ginsburg's illness ...''\}\\ 
\{``ID'': ``N****'', ``title'': ``Ruth Bader Ginsburg defends Kavanaugh, Gorsuch as \\ very decent and very smart'', ``abstract'': ``Supreme Court Justice Ruth Bader\\Ginsburg came to the defense of her more conservative colleagues on the \\ bench, Justices Brett ...'', ``topics'': ``Ruth Bader Ginsburg, Supreme Court \\Justices ...''\}\\ 
\{``ID'': ``N****'', ``title'': ``Ruth Bader Ginsburg back at work after stomach \\bug'', ``abstract'':  ``Supreme Court Justice Ruth Bader Ginsburg returned to \\work Friday morning after missing a day of oral arguments ...'', ``topics'': \\``Ruth Bader Ginsburg ...''\}\\ User Interest:\\ This user is interested in news about{[}politics{]}, especially {[}..., Washington \\ Post, impeachment inquiry, Supreme Court, ...{]}.\end{tabular} \\ \hline
\rowcolor{violet!20}
\multicolumn{1}{c}{\textbf{Output}}                                                                                                                                                                                                                                                                                                                                                                                                                                                                                                                                                                                                                                                                                                                                                                                                                                                                                                                                                                                                                                                                                                                                                                                                          \\ \hline
\rowcolor{violet!2}
\begin{tabular}[c]{@{}l@{}}\{``title'': ``Brett Kavanaugh Gives First Speech as Justice, Praises Ruth Bader \\ Ginsburg Being World Champion'', ``category'': ``politics'', ``topics'': ``Brett \\ Kavanaugh, Ruth Bader Ginsburg, Supreme Court'', ``abstract'': ``In his first \\ speech as a Supreme Court justice, Brett Kavanaugh expressed gratitude \\ to his supporters and hailed Ruth Bader Ginsburg as an inspiration. \\Supreme Court Justice Ginsburg, who recently missed court due to illness,\\ defended Kavanaugh and Gorsuch as 'very decent and very smart.' Despite \\ being absent due to a stomach bug earlier this week, Ginsburg returned \\ to work on Friday.''\}\end{tabular}                                                                                                                                                                                                                                                                                                                                                                                                                                                                                                                                                              \\ \hline

\end{tabular}
\end{table}

\end{document}

%% file: author.tex
\author{Shen Gao}
\authornote{Both authors contributed equally to this research.}
\affiliation{
\institution{University of Electronic Science and Technology of China}
\country{}
}
\email{shengao@pku.edu.cn}

\author{Jiabao Fang}
\authornotemark[1]
\affiliation{
\institution{Shandong University}
\country{}
}
\email{jiabaofang@mail.sdu.edu.cn}

\author{Quan Tu}
\affiliation{
\institution{Renmin University of China}
\country{}
}
\email{quantu@ruc.edu.cn}

\author{Zhitao Yao}
\affiliation{
\institution{Shandong University}
\country{}
}
\email{yaozhitao@mail.sdu.edu.cn}

\author{Zhumin Chen}
\affiliation{
\institution{Shandong University}
\country{}
}
\email{chenzhumin@sdu.edu.cn}

\author{Pengjie Ren}
\affiliation{
\institution{Shandong University}
\country{}
}
\email{jay.ren@outlook.com}

\author{Zhaochun Ren}
\authornote{Corresponding author.}
\affiliation{
\institution{Leiden University}
\country{}
}
\email{z.ren@liacs.leidenuniv.nl}

%% file: intro.tex
\section{Introduction} 

Online news platforms, such as Google News, have become crucial avenues for users to acquire daily information~\cite{Abhinandan2007GoogleNews}. 
However, it is challenging for users to find interesting content among a large number of news articles. 
Hence, the news recommender system, which selects news based on user preference, is designed to improve the experience of user reading and alleviate the information overload problem~\cite{Miaomiao2019Survey}.

\begin{figure}[!t]
\centering
  \includegraphics[width=0.95\columnwidth]{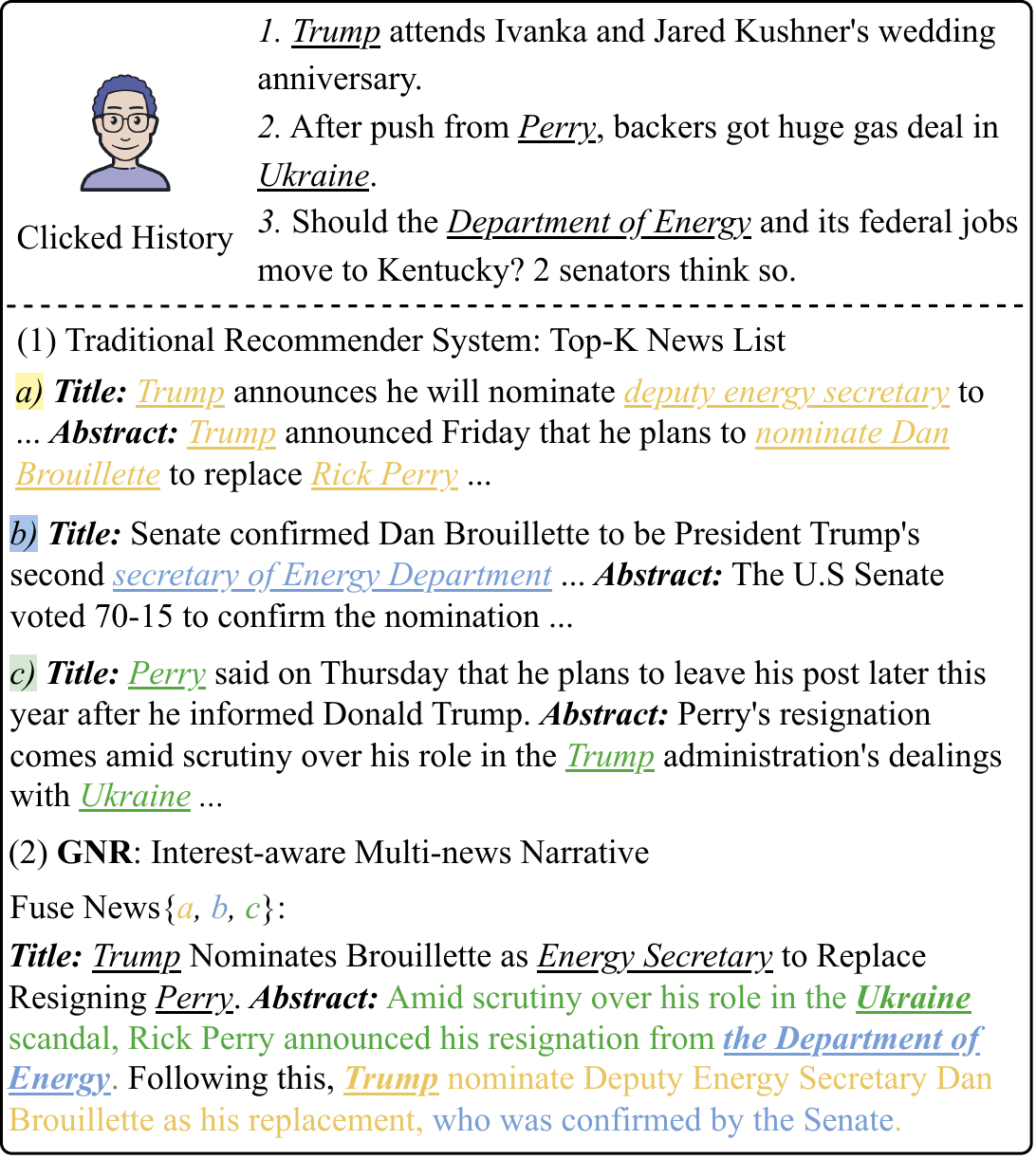}
  \caption{Only recommending existing news in the news corpus is one of the limitations of traditional methods.}
  \label{fig:intro_gen}
\vspace{-4mm}
\end{figure}

Nevertheless, traditional news recommendation encounters the following limitations: 
(1) News recommendation is a content-based task that mainly relies on semantic matching between candidate news and user preference. 
These methods only capture explicit semantic relationships and overlook the equally important implicit relationships required for accurate recommendations.
For example, a news article about ``Argentina’s win over France was the greatest World Cup final ever'' may not exhibit an obvious semantic connection with another news article about ``Lionel Messi cements his place among the greats after winning epic duel against Kylian Mbappé''.
However, a user who likes the previous news is likely to like the latter as well. This is due to their shared theme of ``Messi won the World Cup final'', representing an implicit relationship between them.
However, finding the implicit relationship requires the knowledge of ``Messi is a player of the Argentina national football team'' and reasoning ability.
(2) Existing news recommender systems can only recommend news in its original form. 
Users are required to read numerous lengthy news articles to gain an understanding of the overall context of events. 
Furthermore, users with different interests are presented with identical content without any personalization.
Figure~\ref{fig:intro_gen} shows an example of the recommended news list of existing methods.
Although the recommended news list covers the main events of a user-preferred topic, the user will read several long news articles with redundant information.
A desired outcome for a news recommender system would be to provide a concise paragraph that overviews the main events that the user is interested in.

\begin{figure}[!t]
\centering
  \includegraphics[width=1\columnwidth]{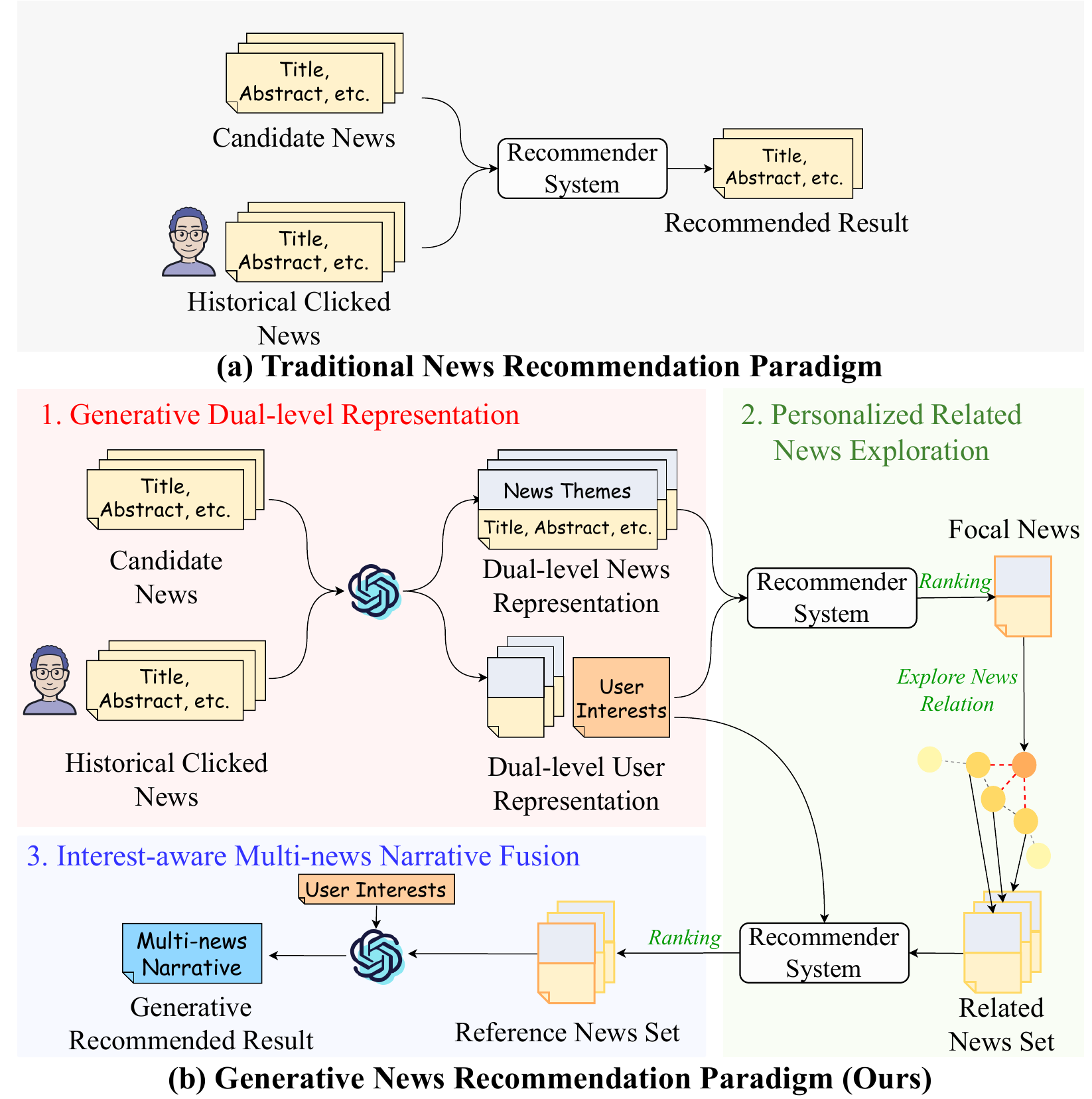}
  \caption{The differences between traditional news recommendation paradigm and our proposed generative news recommendation paradigm \model.}
  \label{fig:intro_overall}
  \vspace{-4mm}
\end{figure}

In this paper, we introduce a novel generative news recommendation paradigm (\model). 
Our approach incorporates the Large Language Model (LLM) as a generator to enhance news recommendations by precisely catering to user needs.
As illustrated in Figure~\ref{fig:intro_overall}(a), traditional news recommendation methods perform semantic matching using candidate news and user representation, primarily composed of the user's historical clicked news list. 
These methods subsequently present news articles to the user in their original format.
In contrast, our approach leverages the LLM to generate theme-level representations for news and users, as depicted in Figure~\ref{fig:intro_overall}(b). 
Then we explore a personalized related news set as the information source and generate a coherent and logically structured narrative to engage users and encourage them to read more of the news. 

Specifically, the \model consists of three modules. 
The first two modules aim at retrieving a news set that contains user-interested news and its related news, and the last module fuses the news in the set into a coherent narrative.
The first module is \textbf{Generative Dual-level Representation}: 
Following the previous news recommendation methods~\cite{Wu2023Survey, wu-etal-2019-neural-news, Wu2021Empowering}, we first obtain semantic-level representations for both users and news. 
Then we leverage the LLM to map news content and user profiles to theme-level representations. 
Finally, we combine these representations into dual-level representations.
The second module is \textbf{Personalized Related News Exploration}: 
To generate coherent narratives, we need to find a personalized and interconnected news set. 
There are three main steps in this module.
We first conduct news ranking based on the dual-level news and user representations and obtain the focal news that best matches user preference.
Second, we propose to explore the logical relation between news articles, aiming to discover more news articles related to the focal news.
Since the second step introduces more related news articles that may not be interesting to the user, we conduct personalized filtering in the third step.
Finally, we obtain a reference news set that encompasses both the main event context of the focal news and takes into account the user preference.
The third module is \textbf{Interest-aware Multi-news Narrative Fusion}: 
The primary objective of this module is to create a coherent and logically structured narrative that encapsulates the central theme of the reference news set. 
To enhance the alignment between the generated narrative and user interests, and to attract users to engage more with the content, we introduce the User Interest Alignment Fine-Tuning (\trainsmall) method, which adjusts the probabilities of multiple multi-news narratives by optimizing for ranking loss.
Extensive experiments conducted on a benchmark dataset demonstrate that \model improves recommendation accuracy and offers users more personalized multi-news narratives. 

To sum up, our contribution can be summarized as follows:

$\bullet$ We propose a generative news recommendation paradigm (\model), which introduces a powerful LLM as the generator to make the recommended news meet user needs more precisely.

$\bullet$ In \model, we design three modules to perform two sub-tasks: (I) Leveraging the internal knowledge and reasoning capability of LLM to retrieve a personalized related news set; (II) Generating a coherent and logically structured narrative, thus engaging users in further reading of the news. 

$\bullet$ We propose a novel training method User Interest Alignment Fine-tuning (\trainsmall) which fine-tunes the LLM through ranking loss based on user interests.

$\bullet$ Extensive experiments on the MIND dataset demonstrate that our \model can significantly improve the accuracy of recommender systems and the generated narratives are more personalized to fulfill the user information needs.

%% file: related.tex
\section{Related Work}
\subsection{Generative LLMs for Recommendation}
Recently, LLMs have achieved great success in many natural language processing tasks due to their excellent natural language understanding and natural language generation abilities~\cite{zhou2023comprehensive,zhao2023survey, sun2023indeterminacy, sun2023convntm, guo2023retrieval}. And many studies have surfaced that LLMs can be used for recommendations due to their strong instruction following and common-sense reasoning abilities~\cite{lyu2023llm,dai2023uncovering,sanner2023large,liao2023llara,mysore2023large,harte2023leveraging}.

Initially, \citet{wang2023zero} proposed leveraging LLMs to conduct the sequential recommendation directly by prompting. 
\citet{hou2023large} proposed using LLMs as a ranker, which converts the interaction history and candidate items into natural language form, and inputs them into LLMs together with the ranking instruction.
However, many researchers~\cite{bao2023bi,bao2023tallrec,zhang2023recommendation} found that when LLMs are directly applied to recommendation tasks through prompting or in-context learning, there is a certain performance gap compared with existing recommendation models. 
The reason is that LLMs are primarily trained on NLP-related tasks and lack training on recommendation tasks~\cite{ouyang2022training,touvron2023llama}.
In order to solve this problem, \citet{bao2023tallrec} and \citet{zhang2023recommendation} proposed to improve the performance of LLMs for recommendation by instruction tuning, which involves generating natural language format instructions based on data from the recommendation tasks. 

In addition to directly using LLMs to conduct recommendation, some researchers propose to use generative LLMs to assist traditional recommender systems~\cite{wang2023large,wang2023enhancing}. 
\citet{gao2023chat} proposed Chat-REC, which leverages an LLM as the controller of the recommender system and the interface with the users. 
Chat-REC allows the users to express their needs proactively and also makes the recommendation more explainable. 
\citet{wang2023rethinking} proposed an interactive evaluation approach based on LLMs to address the problem of past evaluation metrics that overly focus on matching with ``ground-truth'' items in conversational recommender systems.
\citet{wang2023generative} proposed a generative recommender paradigm, which uses LLMs as the controller of the recommender system to determine whether to recommend an existing item from the item corpus or to generate a new item through an AI generator. 
Their method primarily uses LLMs as the controllers of the system and utilizes the diffusion models to transfer the micro-video style. 
Most of the previous works use LLMs as the recommender systems or the controllers. On the contrary, we leverage LLMs as the generator to provide dual-level representations for better recommendation and generate personalized and coherent multi-news narratives that assist users in learning news events.

\subsection{News Recommendation}
The main task of the news recommendation is to recommend news articles that are consistent with the user preferences~\cite{Wu2023Survey,wang2018dkn, yang2023going, li2023exploring}. 
The core of the news recommender system includes the news encoder and the user encoder. 
\citet{Wu2019Neural} mainly focused on enhancing news representation and proposed a news recommendation approach with attentive multi-view learning (NAML), which uses word-level and view-level attention networks to select key information in news. 
\citet{An2019Neural} proposed a neural news recommendation approach with both long-term and short-term user representations. 
\citet{wu2019MHSA} proposed a neural news recommendation approach, which combines the multi-head self-attention mechanism in the news encoder and the user encoder. 
Following previous work,~\citet{Wu2021Empowering} used the more powerful pre-trained language models as the backbone for the news encoder. 
With the rise of LLMs, many researchers have started incorporating it into news recommendation~\citet{runfeng2023lkpnr, li2023preliminary, Liu_Chen_Sakai_Wu_2023,liu2023once}. 
For example, \citet{Liu_Chen_Sakai_Wu_2023} proposed to utilize both open- and closed-source LLMs to enhance content-based recommendation, which includes the news recommendation.

However, these existing news recommendation methods are all based on human-written news corpus, which means they can only recommend raw news articles as-is. In our work, \model can use LLMs to fuse multi-news narratives that more align with user preference.

\begin{table*}[htb]
\caption{Example prompt for theme-level news representation generation}\label{tab:news_prompt}
\small
\renewcommand\arraystretch{1}
\begin{tabular}{>{\centering\arraybackslash}p{15cm}}
\rowcolor{yellow!20}
\hline
\multicolumn{1}{c}{\textbf{Instruction}}                                                                                                                                                                                                                                                                                                              \\ \hline
\rowcolor{yellow!2}
\begin{tabular}[c]{@{}l@{}}Based on the given news information, summarize what \textbf{topic(s)} the news is related to. Each news article is related to 1-3 topics, \\ and each topic should not exceed five words.\end{tabular}                                                                                                                                                \\ \hline
\rowcolor{violet!10}
\multicolumn{1}{c}{\textbf{Input}}                                                                                                                                                                                                                                                                                                                    \\ \hline
\begin{tabular}[c]{@{}l@{}}\{``title'': ``Trump says the Kurds `are no angels' and the PKK are `probably worse' than ISIS'', ``abstract'': ``President Trump defended\\ his decision to withdraw U.S. forces from Syria, claiming that ... and saying that the Kurds `are no angels' ... '', ``category'': ``politics''\}\end{tabular} \\ \hline

\rowcolor{violet!20}
\multicolumn{1}{c}{\textbf{Output}}                                                                                                                                                                                                                                                                                                                   \\ \hline
\rowcolor{violet!2}
This news is related to \textbf{{[}Trump's decision on Syria{]}}, \textbf{{[}Kurds and PKK{]}}.                                                                                                                                                                                                                                                                                          \\ \hline
\end{tabular}
\vspace{-3mm}
\end{table*}

\begin{table*}[htb]
\caption{Example prompt for theme-level user representation generation}\label{tab:user_prompt}
\small
\renewcommand\arraystretch{1}
\begin{tabular}{>{\centering\arraybackslash}p{15cm}}
\rowcolor{yellow!20}
\hline
\multicolumn{1}{c}{\textbf{Instruction}}                                                                                                                                                                                                                                                                                                                                                                                                                                                                                                                                                                                                                                                                                                                                                                                                                                                                                                                                                                                                                                                                                                                                                                                                     \\ \hline
\rowcolor{yellow!2}
\begin{tabular}[c]{@{}l@{}}You are asked to describe user interest based on his/her browsed news list. User interest includes the news \textbf{{[}categories{]}} and news \\ \textbf{{[}topics{]}} (under each \textbf{{[}category{]}}) that users are interested in.\end{tabular}                                                                                                                                                                                                                                                                                                                                                                                                                                                                                                                                                                                                                                                                                                                                                           \\ \hline
\rowcolor{violet!10}
\multicolumn{1}{c}{\textbf{Input}}                                                                                                                                                                                                                                                                                                                                                                                                                                                                                                                                                                                                                                                                                                                                                                                                                                                                                                                                                                                                                                                                                                                                                                                                           \\ \hline

\begin{tabular}[c]{@{}l@{}}News List:\\
\{``ID'': ``News 1'', ``title'': ``Lionel Messi says he wants to ...'', ``category'': ``sports'', ``topics'': ``Argentina football player Lionel Messi''\} \\
\{``ID'': ``News 2'', ``title'': ``How the world reacted to the best World Cup final ever'', ``category'': ``sports'', ``topics'': ``World Cup final'' ...
\\
\end{tabular} \\ \hline
\rowcolor{violet!20}
\multicolumn{1}{c}{\textbf{Output}}                                                                                                                                                                                                                                                                                                                                                                                                                                                                                                                                                                                                                                                                                                                                                                                                                                                                                                                                                                                                                                                                                                                                                                                                          \\ \hline
\rowcolor{violet!2}
\begin{tabular}[c]{@{}l@{}}According to \textbf{{[}News 1, News 2, News 3, News 4{]}}, this user is interested in news about [sports], especially \textbf{{[}Lionel Messi, World} \\ \textbf{Cup final, Argentina's victory in the World Cup{]}}.\end{tabular}                                                                                                                                                                                                                                                                                                                                                                                                                                                                                                                                                              \\ \hline

\end{tabular}
\vspace{-3mm}
\end{table*}

%% file: preliminar.tex
\section{Task Formulation}

The generative news recommendation can be formulated as retrieving a reference news set $\mathcal{N}^r$ from the news corpus $\mathcal{N}$ and generating a coherent narrative $n^m$ to overview the main event of the reference news set $\mathcal{N}^r$.
This reference news set must fulfill two key characteristics. 
Firstly, it should align with the user interests.
Second, it should comprehensively mine related news articles by exploring implicit relationships among news articles, encompassing the full context of an event.
Then the generated multi-news narrative $n^m$ can introduce the full context of the news event.
In this paper, we use $\mathcal{N}=[n_1, n_2, \dots, n_k]$ to denote the whole news corpus which has $k$ news articles in total.
The news recommender system models the user preference based on the user's historical clicked news list $\mathcal{N}^h = [n_1^h, n_2^h, \dots, n_i^h]$. 

Then the recommender system matches the user preference and news in the candidate list $\mathcal{N}^c = [n_1^c, n_2^c, \dots, n_j^c]$ to predict the scores and outputs a focal news $n^f$ with the highest matching scores. Based on the focal news, we apply a filter to the whole news corpus $\mathcal{N}$ in order to find a reference news set $\mathcal{N}^r = [n^f, n_1^r, \dots, n_{T-1}^r]$, which is both personalized and interconnected.
Then we fuse the reference news set $\mathcal{N}^r$ to obtain a multi-news narrative $n^m$ as the generative recommended result.

%% file: method.tex
\section{\model Method}
In order to help news recommendations better satisfy user needs, we propose \fullmodel(\model).

The \model consists of three modules, as shown in Figure~\ref{fig:intro_overall}. 
\textbf{First}, we leverage the LLM to generate theme-level representations and combine them with the semantic-level representations to obtain the dual-level representations, as shown in \S~\ref{sec:dual-level}. 
\textbf{Second}, we design a three-step pipeline for acquiring a personalized and interconnected news set. In this process, we initially rank the candidate news, then explore the relation between the news articles, and finally filter the related news set.
Details are shown in \S~\ref{sec:news-expl}.
\textbf{Third}, we fuse the news set to generate a brief multi-news narrative, which can assist a user in quickly learning about the news event that interests him/her. Details are shown in \S~\ref{sec:news-fus}.

\subsection{Generative Dual-level Representation}\label{sec:dual-level}
Higher-level connections can help news recommender systems better match users and news, and such connections require domain knowledge and reasoning ability to be obtained. Therefore, we propose to use the LLM to generate theme-level representations for both news and users.
Finally, we combine the theme-level representations with the semantic-level representations to obtain dual-level representations and promote more accurate matching. 

\subsubsection{Theme-level News Representation}

To obtain higher-level news representations, we leverage the common-sense knowledge in the LLM to summarize the themes for each news.
For example, the original content of a news article is ``In the 2022 FIFA World Cup, Lionel Messi cements his place among the greats after winning epic duel against Kylian Mbappé''. 
Then, the theme of this news is ``Messi won the World Cup'', which also serves as the theme-level representation of this news.
Specifically, we first manually construct a prompt template, and then put the original news content as input, including news titles, abstracts, and categories. 
The specific prompt construction is shown in Table~\ref{tab:news_prompt}.

\subsubsection{Theme-level User Representation}

We consider user profiles, which encompass various news themes, as higher-level representations of users. To obtain this representation, we employ the LLM to infer the connections within each news in the user's historical clicked list and generate a description of the user profile.
For example, if a user's historical clicked news list contains ``Why Argentina’s win over France was the greatest World Cup final ever'' and ``Lionel Messi cements his place among the greats after winning epic duel against Kylian Mbappé'', we can infer that this user is interested in the news theme ``Argentina won the World Cup'', which is also part of the theme-level user representation.
Furthermore, we leverage the LLM through in-context learning. We manually create a prompt template and input news information, which includes news titles, categories, and theme-level news representations.
The specific prompt construction is shown in Table~\ref{tab:user_prompt}.

\subsubsection{Dual-level Representation Combination}

As mentioned above, we leverage the LLM to generate theme-level representations for news and users. 
Meanwhile, we can get the semantic-level representation based on the original content of the news. 
Then we need to combine these two representations and provide the dual-level representations to the recommender system. 
Inspired by NAML~\cite{Wu2019Neural}, which is a widely known news recommendation method, we also use the multi-view attention network to fuse the representations:
\begin{equation}
\begin{gathered}
\alpha_s=q_{v1}^T \tanh \left(\operatorname{Linear}\left(e_s\right)\right), \alpha_t=q_{v2}^T \tanh \left(\operatorname{Linear}\left(e_t\right)\right) \\
\alpha_s^{\prime}=\frac{\exp \left(\alpha_s\right)}{\exp \left(\alpha_s\right)+\exp \left(\alpha_{\mathrm{t}}\right)}, \alpha_t^{\prime}=\frac{\exp \left(\alpha_t\right)}{\exp \left(\alpha_s\right)+\exp \left(\alpha_{\mathrm{t}}\right)} \\
e_d=\alpha_s^{\prime} e_s+\alpha_t^{\prime} e_t,
\end{gathered}
\end{equation}
where $q_{v_1}$ and $q_{v_2}$ both are attention query vectors, $e_s$ is the embedding of semantic-level representation, $e_t$ is the embedding of theme-level representation, $e_d$ is the embedding of dual-level representation, $\alpha_s$ is the attention weight of the semantic-level representation, and $\alpha_t$ is the attention weight of the theme-level representation.
In the following, we default the user embedding and the news embedding mentioned in the following are calculated based on the dual-level representations.

\subsection{Personalized Related News Exploration}\label{sec:news-expl}

To ensure the coherence of the recommended narrative and cater to user interests, it is imperative to extract a personalized and  interconnected news set, referred to as a ``reference news set'', from the news corpus $\mathcal{N}$. 
To achieve this goal, we have devised a three-step pipeline, which encompasses dual-level news ranking, news relation exploration, and personalized filtering. 
In the subsequent sections, we introduce the details of these three steps.

\subsubsection{Dual-level News Ranking}\label{sec:news-ranking}

To cater to user interests effectively, we adopt a candidate news ranking method similar to the conventional news recommendation paradigm. 
To train the ranking model, we use the negative sampling and randomly select $K_{neg}$ non-clicked candidate news as negative samples.
Then the probability of the user clicking the positive candidate news is:
\begin{equation}
\begin{gathered}
\hat{y}_i = e_i^{user} \cdot e_i^{cand}, \\
p_i = \frac{\exp \left(\hat{y}_i^{+}\right)}{\exp \left({\hat{y}_i^{+}}\right)+\sum_{j=1}^{K_{neg}} \exp \left(\hat{y}_i^j\right)}
\end{gathered}
\end{equation}
where $e_i^{user}$ is the embedding of dual-level user representations, $e_i^{cand}$ is the embedding of dual-level candidate news representations and $\hat{y}_i$ is the predicted ranking score of the candidate news.

We optimize the probability of positive sample $p_i$ through log-likelihood loss:
\begin{gather}
    \mathcal{L}_{ranking} = -\sum_{i \in S} \ln \left(p_i\right)
\end{gather}
where $S$ is the training dataset.

Then, we can obtain a ranking score for each candidate news, indicating the matching degree between the candidate news and user preference. 
However, the top news articles in the ranking list may not be correlated with each other, making it challenging to generate a coherent narrative based on these news articles.
Thus, we only select the top-1 news as the focal news $n^f$ in the ranking list and use it to conduct the subsequent steps. 

\subsubsection{News Relation Exploration}\label{sec:rela-exp}
After obtaining the focal news $n^f$, we need to find some related news articles from the news corpus to generate a coherent narrative. 
Therefore, we implement a news relation classifier to judge whether two news articles are related and set a relation threshold $\alpha$.
In order to train the model, we collected a set of related news pairs from news websites to form a training dataset. 
We construct the news relation classifier based on Siamese networks~\cite{reimers2019sentence} and train the model using contrastive learning loss. 
The positive pair is a pair of news articles that are related to each other, and we randomly select some unrelated news articles to construct negative data pairs. 
The loss function is formulated as follows:
\begin{gather}
\mathcal{L}_{classify}=\max \left(\left\|e_i^{news}-e_j^{news}\right\|-\left\|e_i^{news}-e_k^{news}\right\|+\epsilon, 0\right)
\end{gather}
where $e_j^{news}$ is the embedding of the positive news (\aka related news), $e_k^{news}$ is the embedding of the negative news (\aka unrelated news), and $\epsilon$ is the margin between positive and negative pairs.
After model training, we use this news relation classifier to explore a related news set $\mathcal{N}^{rel} = \{n^{rel}_1, n^{rel}_2, \dots, n^{rel}_j\}$ related to the focal news $n^f$ from the news corpus $\mathcal{N}$.

\subsubsection{Personalized Filtering}

Since the related news set $\mathcal{N}^{rel}$ is explored based on the news relation, which ignores the user preference.
In this module, we propose to personalized filter the related news set using the traditional recommender system.
We use the related news set $\mathcal{N}^{rel}$ as the candidate set and compute the matching score $\hat{y}_i$ between related news embedding $e^{rel}_i$ and user embedding $e^{user}_i$ as follows:
\begin{gather}
\hat{y}_i=e_i^{user} \cdot e_i^{rel}
\end{gather}
Then we select $T-1$ news articles with the highest matching score. And together with the focal news $n^f$ obtained above, we form the reference news set $\mathcal{N}^r$.

\begin{figure}[t]
\centering
  \includegraphics[width=1\columnwidth]{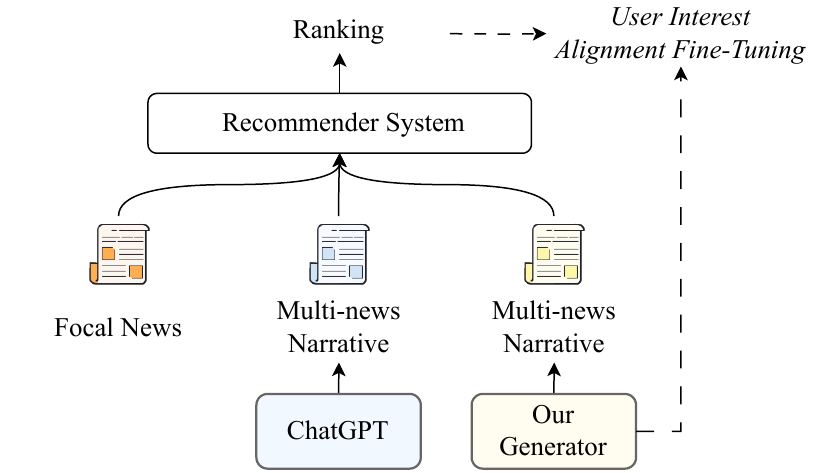}
  \caption{The framework of UIFT method.}
  \label{fig:uift}
\vspace{-4mm}
\end{figure}

\subsection{Interest-aware Multi-news Narrative Fusion}\label{sec:news-fus}

Conventional news recommendation methods typically recommend news articles in their original form. 
Consequently, when users express interest in a specific news event, they need to read a multitude of related news articles to gain an understanding of the pertinent content. 
This process is notably time-consuming and results in a suboptimal user experience. 
Simultaneously, personalization diminishes as recommender systems deliver the same news content to users with varying interests. 
To address this challenge, we propose the fusion of multi-news narratives based on user interests. 
In this section, we first introduce the process of multi-news narrative fusion and subsequently introduce the \trainsmall method, designed to tailor the generated narrative more closely to user interests.

First, to ensure the coherence and readability of the narrative, we use the focal news $n^f$ (as introduced in \S~\ref{sec:news-ranking}) as the central point of the multi-news narrative. 
Then the goal of the generated narrative is to extract the key facts of the news set $\mathcal{N}^r$ that align with user interests and fuse these key facts around the focal news $n^f$.
An illustrative example can be found in \S~\ref{sec:fuse_prompt}.

Despite the in-context learning capabilities of general black-box LLM, they still exhibit deficiencies in capturing user interests, a crucial aspect of the recommendation task. 
So we propose User Interest Alignment Fine-Tuning (referred to as \trainsmall), aimed at enhancing the personalization of LLM-generated multi-news narratives. 
The framework of \trainsmall is shown in Figure~\ref{fig:uift}.
We first utilize ChatGPT to generate the multi-news narrative and conduct supervised fine-tuning of our multi-news narrative generator to distill the ability and knowledge from ChatGPT to our narrative generator.
Then \trainsmall trains our multi-news narrative generator by incorporating ranking loss, thereby aligning its ranking for multiple news narratives with user interests.
We collect a training set containing multiple focal news articles $n^f$, the corresponding multi-news narratives generated by ChatGPT $n^{ChatGPT}$, and the corresponding multi-news narratives fused by our generator $n^{\model}$. 
Next, \trainsmall first allows the narrative generator $\pi$ trained with supervised fine-tuning to predict the conditional probability $p_i$ for each news:
\begin{gather}
p_i = \frac{\sum_t \log \mathrm{P}_\pi \left(s_{i, t} \mid n, s_{i<t}\right)}{\left\|s_i\right\|},
\end{gather}  
where $n$ is the news in the set $\{n^{ChatGPT}_i, n^{\model}_i, n^f_i\}$.

In \trainsmall, the model learns to give higher probabilities to more personalized news, thus aligning with user interests. 
Specifically, we rank the news set $\{n^{ChatGPT}_i, n^{\model}_i, n^f_i\}$ based on the user interests. 
Due to the high cost of manual annotation, we rely on a well-trained news recommender system to reflect user interests.
Based on the predicted score computed by the recommender system, we can get the ranking between the news (i.e., $r^{\model}_i<r^{ChatGPT}_i<r^f_i$) and the $r_i \in \{1,2,3\}$ denotes the ranking of the news. When $r_i$ is smaller, it means that the corresponding news is more aligned with the user interests.

Our training goal is to give the larger probability $p_i$ to news with better ranking $r_i$.
We achieve this through ranking loss:

\begin{gather}
L_{\text{rank}}=\sum_{r_i<r_j} \max \left(0, p_i-p_j\right),  
\end{gather}

%% file: exp.tex
\section{Experiment}
In this section, we conduct extensive experiments to answer the following research questions:

\textbf{RQ1}: How much accuracy improvement can \model bring to the recommendation models by combining dual-level representations?

\textbf{RQ2:} Can \model generate personalized and factually consistent multi-news narratives?

\textbf{RQ3}: When exploring the relation between news, how does the relation threshold $\alpha$ affect the performance?

\textbf{RQ4}: When retrieving the reference news set $\mathcal{N}^r$, how does the maximum number of reference news $T_{max}$ affect the performance?

\textbf{RQ5}: What are the differences between personalized multi-news narratives and non-personalized multi-news narratives?

\subsection{Datasets}
We conduct experiments on the MIND dataset~\cite{Wu2020MIND}, which is a large-scale dataset for news recommendation. 
MIND dataset contains the news dataset and the behaviors dataset.
Each news item in MIND contains a news title, abstract, etc.
Each behavior includes a historical clicked news list $\mathcal{N}^h$ and a candidate news list $\mathcal{N}^c$.
For the news dataset, we first filtered 5145 news articles from MIND-Large under the ``politics'' category, which is more appropriate for the scenario of \model. 
We will extend the \model to experiment under more news categories in the future. 
For the behaviors dataset, we first filtered behaviors to ensure all news items in the historical clicked list and the candidate list belong to the politics news dataset. 
When fusing multi-news narratives, we filter behaviors to ensure that the length of the historical clicked news list to fall within the range: $5<=|\mathcal{H}|<=15$.
We then structured two separate datasets for training the recommender system and the multi-news narrative generator respectively. 
The sizes of the training, validation, and test sets for the recommender system are 43232, 4800, and 6713. 
Similarly, for the multi-news narrative generator, the sizes of the training, validation, and test sets are 8926, 183, and 1956.

Furthermore, to train the classifier for news relation exploration, we employed a web crawler to extract news articles from the CNN website~\footnote{https://edition.cnn.com/}. We also scraped the relevant news from each news webpage to construct the training dataset for the news relation classifier. While retrieving the reference news, we used them to augment the related news and enrich our news dataset.

\begin{table*}[htb]
\caption{Performance of recommendation accuracy. We experiment with different combinations of news representations and user representations. ``Sem'' denotes semantic-level representation. ``Dual'' denotes dual-level representation.}\label{tab:recommend}
\renewcommand\arraystretch{1.1}
\normalsize
\begin{tabular}{cc>{\centering\arraybackslash}p{0.2cm}>{\centering\arraybackslash}p{1.2cm}>{\centering\arraybackslash}p{1.2cm}>{\centering\arraybackslash}p{1.2cm}>{\centering\arraybackslash}p{0.2cm}>{\centering\arraybackslash}p{1.2cm}>{\centering\arraybackslash}p{1.2cm}>{\centering\arraybackslash}p{1.2cm}>{\centering\arraybackslash}p{0.2cm}>{\centering\arraybackslash}p{1.2cm}>{\centering\arraybackslash}p{1.2cm}>{\centering\arraybackslash}p{1.2cm}}
\toprule
\multicolumn{2}{c}{\begin{tabular}[c]{@{}c@{}}Input Type\end{tabular}} &  & \multicolumn{3}{c}{NRMS}                         &           & \multicolumn{3}{c}{PLM4NR (title)}                  &           & \multicolumn{3}{c}{PLM4NR (title and abstract)}        \\ \cline{1-2} \cline{4-6} \cline{8-10} \cline{12-14} 
News                                     & User                                    &  & NDCG@5         & AUC            & MRR            &           & NDCG@5         & AUC            & MRR            &           & NDCG@5         & AUC            & MRR            \\ \midrule
Sem                                      & Sem                                     &  & 58.57          & 56.47          & 50.44          &           & 62.38          & 69.44          & 54.73          &           & 61.11          & 68.36          & 53.70          \\
Sem                                      & Dual                                   &  & \underline{59.39}          & \underline{56.54}    & 51.36          &           & 62.98          & \textbf{70.38} & 55.21          &           & 61.78          & 68.77          & 54.35          \\
Dual                                    & Sem                                     &  & 58.99    & 56.35          & \underline{51.46}    &           & {\underline {63.02}}    & 69.95          & {\underline {55.61}}    &           & {\underline {62.76}}    & {\underline {69.26}}    & {\underline {55.52}}    \\
Dual                                    & Dual                                   &  & \textbf{59.81} & \textbf{57.54} & \textbf{51.80} & \textbf{} & \textbf{63.46} & {\underline {70.31}}    & \textbf{56.03} & \textbf{} & \textbf{62.99} & \textbf{69.42} & \textbf{55.72} \\ \bottomrule
\end{tabular}
\vspace{-1mm}
\end{table*}

\subsection{Evaluation Metrics}
The \model proposed two sub-tasks: (1) Retrieving personalized references news sets; (2) Fusing coherent multi-news narratives. We evaluate the performance of these two sub-tasks separately.

For the first sub-task, we evaluate whether the theme-level representations generated by \model can promote recommendation accuracy. So we leverage AUC (Area Under the Curve), MRR (Mean Reciprocal Rank), and NDCG@K (Normalized Discounted Cumulative Gain) where K=5 as the evaluation metrics. 

For the second sub-task, as \model is a novel paradigm, there are no previous benchmarks available. Therefore, we propose two automatic metrics to evaluate the personalization and consistency of multi-news narratives: 
(1) \textbf{Win Rate} evaluates whether the fused multi-news narratives are more personalized than the corresponding focal news. 
We first calculate the predicted scores of the multi-news narrative and the focal news based on a well-trained news recommendation model.
The inputs of the recommender system are the dual-level news representations and user representations. 
We mark a situation as a ``Win'' when the predicted score of a multi-news narrative surpasses that of the focal news, interpreting this as an indication that the multi-news narrative aligns better with the user preference. 
Then we compute the Win Rate across the entire test dataset.
(2) \textbf{Consistency Rate} is calculated between the reference news sets and the multi-news narratives. During the evaluation, we feed a reference news set and a multi-news narrative to the evaluator. The evaluator then determines whether the reference news set and the multi-news narrative are consistent. Subsequently, we compute the consistency rate across the entire test dataset.
And inspired by \citet{luo2023chatgpt}, we use the ChatGPT (gpt-3.5-turbo)~\footnote{https://chat.openai.com} as the evaluator to determine consistency.

\subsection{Baseline Models}
We evaluate \model against the following news recommendation methods:
\textbf{(1) NRMS~\cite{wu-etal-2019-neural-news}} leverages the multi-head self-attention to learn news representations and capture the relatedness between the news;
\textbf{(2) PLM4NR (title)~\cite{Wu2021Empowering}} uses pre-trained language models to model news representations from news titles. We leverage the best variant PLM4NR-NRMS in our experiments;
\textbf{(3) PLM4NR (title and abstract)} is similar to PLM4NR (title), but it models news representations from titles and abstracts, not just titles.

\subsection{Implementation Details}
For theme-level representation generation, we select the ChatGPT (gpt-3.5-turbo) as the backbone model.
For news relation exploration, we use the SBERT~\cite{reimers2019sentence} as the backbone model. During the training, we use AdamW as the optimizer, and the learning rate is set to 1e-5. 
When selecting the related news, we set hyperparameter $\alpha=0.8$.
During the training of traditional news recommender systems, the learning rate is set to 1e-4.
In the PLM4NR model, we employ DistilBERT as the encoders and utilize the output embedding of the ``[CLS]'' token.
Meanwhile, we treat the next news to be clicked (i.e., ground truth) as the focal news to better compare whether the multi-news narratives fused by \model can better match user preferences.
We set the length of reference news list $T$ to fall within the range: $2<|T|<5$.
We exclude samples that do not have a sufficient number of related news articles, as they are not suitable for our method.
For our multi-news narratives generator, we use the LLaMA 7B as the backbone model.
And for RQ3, RQ4, and RQ5, we utilize ChatGPT as the generator for a fair comparison.
During the supervised fine-tuning of LLaMA, the learning rate is set to 5e-5. 
During the \trainsmall, the learning rate is set to 1e-5. 
Besides, all LLaMA-based experiments are conducted on 8 80GB Nvidia A800 GPUs, and other experiments are conducted on 24G Nvidia 3090 GPUs.

\subsection{Performance of Recommendation Accuracy (RQ1)}
We first evaluate the recommendation accuracy when using the dual-level representations as input, and report the results in Table~\ref{tab:recommend}.
From the results, we can have the following observations.

\subsubsection{Result Analysis} 
Compared to the accuracy of using the semantic-level representation only, the performance of the backbone models increases significantly when using the dual-level news and user representations. For example, the PLM4NR (title) model improved from 62.38 to 63.46 on metric NDCG@5 and improved from 69.44 to 70.31 on metric AUC.
This suggests that with the help of common-sense knowledge in the LLM, the news recommender systems can capture more levels of relationships between the news and user preference. These relationships can help systems better match the candidate news and users and improve the recommendation accuracy.

\subsubsection{Ablation Study} 
To separately evaluate the effect of the theme-level news representation and the theme-level user representation, we conducted ablation experiments with two settings: (1) using dual-level news representations and semantic-level user representations; (2) using semantic-level user representations and dual-level news representations.
As observed, the recommendation models outperform the baseline models when utilizing theme-level news (or user) representations. 
However, their performance is not as good as the models that employ both theme-level news representations and user representations.
So these results illustrate that both theme-level news representations and theme-level user representations can promote the performance of the recommendation models.

\begin{table}[!t]
\caption{Performance of multi-news narrative fusion. Win Rate is evaluated by PLM4NR (title + abstract).}\label{tab:generation}
\renewcommand\arraystretch{1.1}
\normalsize
\begin{tabular}{>{\centering\arraybackslash}p{1.8cm}>{\centering\arraybackslash}p{1.5cm}>{\centering\arraybackslash}p{2.5cm}}
\toprule
Generator                        & Win Rate & Consistency Rate \\ \midrule
ChatGPT               & 72.80    & 96.63       \\
Ours (SFT)                        & 65.13    & 96.52       \\
Ours (\trainsmall) & 74.54    & 96.57       \\ \bottomrule
\end{tabular}
\vspace{-3mm}
\end{table}

\subsection{Performance of Multi-news Narrative Fusion (RQ2)}
In this part, we evaluate whether the multi-news narratives fused by \model can perform better than the focal news and whether the multi-news narratives are factually consistent with the corresponding reference news. The results are reported in Table~\ref{tab:generation}.

When using ChatGPT as the generator, \model has great performance in both Win Rate and Consistency Rate. It illustrates that ChatGPT is able to fuse coherent and personalized multi-news narratives. We consider this to be due to ChatGPT's excellent in-context learning capability, allowing it to perform our desired task without additional fine-tuning.

When our narrative generator is trained only through supervised fine-tuning, it can effectively fuse news and generate coherent narratives. Nevertheless, its performance in Win Rate is considerably less effective than that of ChatGPT. However, when our narrative generator is trained through \trainsmall, the fused multi-news narratives can achieve superior personalization while maintaining good consistency. We attribute this to the fact that we align the probabilities of our narrative generator with user interests via ranking loss. This alignment aids the model in better understanding user preference during the process of multi-news narrative fusion.

\subsection{Threshold of News Relation (RQ3)}
As described in \S~\ref{sec:news-expl}, we design to retrieve a reference news set containing $T$ personalized and related news. Therefore, we plan to explore the news relation and find a news set related to the focal news.
In this process, we define a relation threshold $\alpha$ to determine whether two news articles are related. 
We think this threshold $\alpha$ can influence the performance of multi-news narrative fusion.
To evaluate the impact, we conduct separate experiments by setting different thresholds. 
To highlight the relation threshold impact, we avoid selecting the same samples for different thresholds. 
During the experiment with threshold $\alpha_1$, certain samples are excluded if the similarity scores between their focal news and no less than $T-1$ news articles surpass the threshold $\alpha_2$, where $\alpha_2 > \alpha_1$.
Then we select 100 test samples for each threshold separately.
However, there are only 83 test samples that comply with $\alpha=0.6$, so we select all of them.
Finally, we evaluate the Consistency Rate of the multi-news narratives under each setting, and the results are shown in Table~\ref{tab:relationship_threshold}.

From the results, we can see that relation threshold $\alpha$ has an impact on the consistency of the fused narrative. We hypothesize that this is due to the fact that when the relevance of the reference news set is low, the generator is unable to reason correctly about the associations between the reference news, leading to the hallucination generation.

\begin{table}[htb]
\caption{The impact of relation threshold $\alpha$ on multi-news narratives.}\label{tab:relationship_threshold}
\normalsize
\renewcommand\arraystretch{1.1}
\begin{tabular}{>{\centering\arraybackslash}p{3.5cm}>{\centering\arraybackslash}p{3.5cm}}
\toprule
Relation Threshold $\alpha$                            & Consistency Rate \\ \midrule
0.6          & 66             \\
0.7         & 85             \\
0.8                      & 98          \\
\bottomrule
\end{tabular}
\vspace{-2mm}
\end{table}

\subsection{The Maximum Number of Reference News (RQ4)}
In this part, we hypothesize that the maximum number of reference news $T_{max}$ determines the quality of input information, thus affecting the final generation quality. Therefore, we experiment with how the maximum number $T_{max}$ promotes or reduces the performance of fusion. 
Considering the limitation of the LLM input length, we set $T_{max}$=2,3,4,5,6 respectively to conduct experiments and evaluate the fused narratives in each setting.

\begin{figure}[!t]
\centering
  \includegraphics[width=0.75\columnwidth]{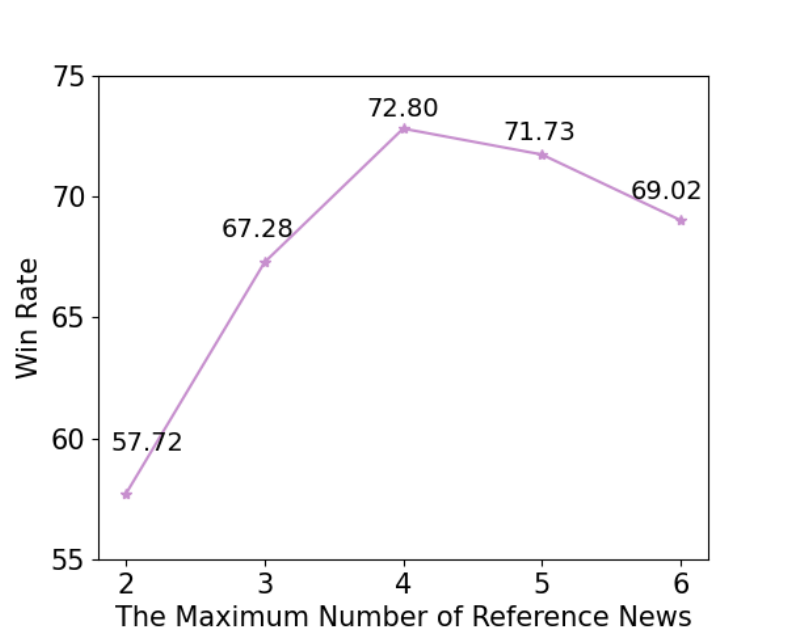}
  \caption{The impact of the maximum number of reference news $T_{max}$ on multi-news narratives.}
  \label{fig:RefNumber}
  \vspace{-2mm}
\end{figure}

As we can see in Figure~\ref{fig:RefNumber}, the Win Rate increases first and then decreases with an increase in the maximum number of reference news $T_{max}$. The optimal performance is achieved when $T_{max}=4$. 
This observation demonstrates when $T_{max}$ is less than 4, the reference news set contains insufficient information to adequately cover the main events and cater to user preference. 
When $T_{max}$ exceeds 4, there is an abundance of information within the reference news set that may lack relevance to the user interests, thereby introducing noise during the fusion process.

\begin{table}[!t]
\centering
\caption{The comparison between non-personalized multi-news narratives and personalized multi-news narratives. The Cohen's kappa between two results is more than $0.4$.}
\renewcommand\arraystretch{1.05}
\label{table:personalized_eval}
\begin{tabular}{@{} cccc @{}}
\toprule
Evaluator & \makecell[c]{Non-Personalized \\ Narrative Wins} & Tie & \makecell[c]{Personalized \\ Narrative Wins} \\

\midrule

Human & $4\%$  & $53\%$  & $43\%$ \\
GPT-4 & $7\%$  & $56\%$  & $37\%$ \\

\bottomrule
\end{tabular}
\vspace{-3mm}
\end{table}

\subsection{Personalized Evaluation of Multi-news Narratives (RQ5)}
In \model, personalized multi-news narratives should prioritize content that aligns with user preference while minimizing irrelevant material. In this section, we provide a quantitative comparison between personalized and non-personalized multi-news narratives. We utilize the same LLM to summarize the reference news sets and obtain non-personalized multi-news narratives. 
In the experiment, we sample 100 test cases and conduct both GPT-4 evaluation and human evaluation. The evaluator should select narratives that highlight content engaging with the user interests while excluding content that doesn't align with those interests.
As shown in Table~\ref{table:personalized_eval}, the results demonstrate that our personalized narratives are more aligned with the user preference.

%% file: conclusion.tex
\section{Conclusion}
In this paper, we introduce a novel generative news recommendation paradigm (\model), which aims at enhancing news recommendation and fulfilling user needs more precisely using LLM. 
By harnessing the internal knowledge and reasoning capabilities of LLM, we generate theme-level representations for news and users. 
These representations are then combined with semantic-level representations to create dual-level representations. 
Subsequently, we explore the news relation to find personalized related news sets based on dual-level representations. 
Afterward, we fuse the personalized related news sets to create coherent and logically structured multi-news narratives, engaging users further in further reading of the news. 
Extensive experiments demonstrate that our \model enhances recommendation performance and generates personalized and factually consistent multi-news narratives.